\documentclass[aip,apl,preprint]{revtex4-1}
\usepackage{amsfonts}
\usepackage{amsmath}
\usepackage{amssymb}
\usepackage{graphicx}

\newcommand{\rvec}{\mathbf{r}}

\newcommand{\Kvec}{\mathbf{K}}

\begin{document}
\title{Robust one-dimensional wires in lattice mismatched bilayer graphene}
\author{Anthony R. Wright}
\affiliation{Institute for Theoretical Physics, University of Leipzig, D-04103 Leipzig, Germany}
\author{Timo Hyart}
\affiliation{Institute for Theoretical Physics, University of Leipzig, D-04103 Leipzig, Germany}
\begin{abstract}
We show that lattice mismatched bilayer graphene can realize robust one-dimensional wires. By considering a
single domain wall where the masses of the Dirac electrons change
their sign, we establish a general projection principle. This
determines how the existence of topological zero-energy domain wall
states depends on the direction of the domain wall and locations of
the massive Dirac cones inside the bulk Brillouin zone. We generalize
this idea for arbitrary patterns of domain walls, showing that the
topologically protected states exist only in the presence
of an odd number of topological domain walls. \end{abstract}

\maketitle

Two of the most prominent topological phases of matter in condensed matter physics occur in the quantum Hall effects \cite{TKNN, qhrev} and topological insulators \cite{km, bz}. The former exhibits a pure topological order \cite{wen} that relies on no
unbroken symmetries, while the latter relies on the retention of time
reversal symmetry. Although a true topological insulator state in graphene is difficult
to observe because of its weak spin-orbit coupling strength, it is
well-established that the massive Dirac cones form topological
structures -- merons -- in momentum space \cite{bz}. Therefore
topologically protected states can still emerge at the domain walls
and edges of gapped graphene and bilayer graphene \cite{semenoff, soliton, blg1,blg2, loss, toprev}. However,
these states are yet to be realised in the lab.

The purpose of the current work is two-fold. Firstly, we propose an ideal material for experimentally realizing one-dimensional wires as topological domain walls: mismatched substrate or bilayer graphene. The canonical example of lattice-mismatched graphene is a twisted bilayer. This material is `trivial' in the context of our results. Two `topological' examples are bilayer graphene where one layer has a single line defect \cite{linedefect}, or where one layer is uni-axially strained. The most topical example of mismatched substrates at present, is graphene atop hexagonal boron nitride \cite{BN}.

\begin{figure}[tbp]
\centering\includegraphics[width=8cm]{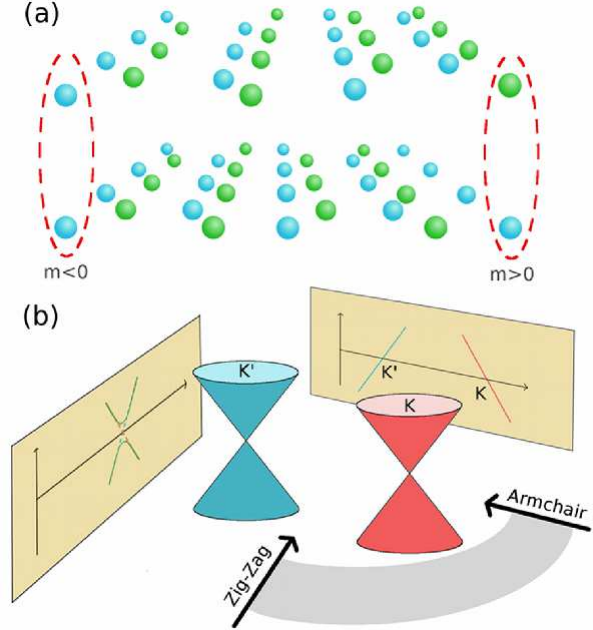}
\caption{The two necessary ingredients to realize robust one-dimensional wires. (a) Spatially varying mass: in mismatched bilayers, interlayer coupling maps onto an effective Dirac mass. (b) The generalized projection principle: in the particular case domain walls in graphene, the projected cones overlap and so may become gapped in the armchair case, whereas for zig-zag walls they remain distinct.}
\label{figmismatch}
\end{figure}

The parallel purpose of this work is to understand the general
properties of multiple domain walls in samples where the
phase-relaxation length is large in comparison to the length scale of
the domain wall patterns. We find that the prescription of the
conditions under which robust zero-energy domain wall states will
exist, contains two ingredients. First, we establish a general
projection principle, which determines how the existence of
topological zero-energy domain wall states depends on the direction of
the domain wall and locations of the massive Dirac cones inside the
bulk Brillouin zone. Then, we generalize this idea for arbitrary
patterns of domain walls and show that the topologically protected
zero-energy states exist only in the presence of an odd number of
topological domain walls, resulting in a novel type of even-odd
effect.

It is well known that at the interface between two massive Dirac systems with opposite signs of mass, which we shall call a domain wall, a zero-energy localized state exists. This observation forms the basis for the existence of surface states in topological insulators, and has also been pointed out in the context of gapped graphene \cite{semenoff}. Creating an oscillating mass term is not trivial. In this work we induce a mass on an effective single layer Dirac system by spatially varying the inter-layer coupling in a bilayer system. The connection between on-site energy and Dirac mass has been established elsewhere \cite{blgmoire}, and the principle is outlined in Fig. \ref{figmismatch}(a). We assume that the states with different spins do not couple. The effective theory of a single spin species system with $N$ massive Dirac cones is

\begin{equation}
H(k) = \sum_i^N \mathbf{d_i}\cdot \vec{\sigma}_i
\label{H0}
\end{equation}
where, for example in graphene there are two cones described by $\mathbf{d_{1(2)}} = (\pm v_F(k_x - K^{1(2)}_x), v_F(k_y-K^{1(2)}_y),m)$, where $v_F = 3t/2$ is the Fermi velocity. and $m$ is a constant mass. The low energy theory for spatially varying mass is obtained with the replacements $m \rightarrow m(\rvec)$, and $k_{x(y)} \rightarrow -i\partial_{x(y)}$.

Firstly, we introduce a Ôprojection principleÕ which defines the
topological nature of the domain wall states for a bulk material with
any number of Dirac cones. For simplicity, we consider first a single
domain wall and assume that the low-energy spectrum of the bulk
material is described with two Dirac cones as shown in
Fig. \ref{figmismatch}(b).  Because of the
translational invariance of the system along the domain wall, the
component of k in that direction remains a good quantum number and the
states in the 2D Brillouin zone of the bulk material become
effectively projected onto a 1D Brillouin zone, potentially causing
mixing of the states with the same value of k along the domain wall
direction.  Therefore, we can expect two possible regimes to arise
from the projections depending on the domain wall directions, as
schematically illustrated in Fig. \ref{figmismatch} (b). The first regime, which we
shall call the topological regime, is that where the low-energy parts
of the two cones are projected onto different regions in the 1D
Brillouin zone. The second regime, which we call the trivial regime,
is that where the projection causes the two valleys to overlap at low
energies. It turns out that this distinction between topological and
trivial regimes is necessary to determine the existence of robust
zero-energy eigenstates localised on the domain wall. Similarly, for a
multiple Dirac cone bulk system, the domain wall states caused by a
given Dirac cone are topologically protected if the projected Dirac
cone retains the distinction from any of the other projected Dirac
cones.

Algebraically applying this principle to Hamiltonian (\ref{H0}), we consider a domain wall at $y=0$, and look for solutions where $E = 0$, obtaining $\psi^T_{k_x}(x,y) = \exp[ik_x x + iK_y y -v_F^{-1}\int_0^y \mathrm{d} y' m(y')](1\,\,-1)$ per cone. If the two cones are at positions $\Kvec$ and $\Kvec'$, as in graphene, then the two projected solutions are separated by $K_x - K_x'$. In the trivial regime however, the two low energy valleys overlap, and admixing of the zero energy solutions occurs, analogous to the case of an armchair ribbon \cite{bf}. In this case then, the zero energy solutions split into two particle-hole symmetric finite energy solutions.

Projecting these solutions onto a $1D$ domain wall parallel to $\hat{d}$, we obtain the effective theory for an $N$ cone bulk system
\begin{equation}
H(k) =  \sum_i^N s_i(k - \mathbf{K^i}\cdot\hat{d})
\label{hlow}
\end{equation}
where $s_i = \pm1$, depending on the corresponding $\mathbf{d}_i$ in the Hamiltonian (\ref{H0}). This is a generalization of the field theory proposed in reference \cite{semenoff}, and is simply a Luttinger liquid. Importantly, in the absence of scattering of magnitude $|(\Kvec^i-\Kvec^j)\cdot\hat{d}|$ for all $j$, we have a single distinct zero-energy surface solution arising from cone $i$.

\begin{figure}[tbp]
\centering\includegraphics[width=8.5cm]{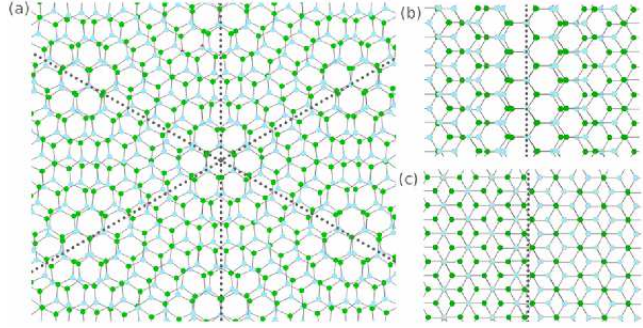}
\caption{Three mismatched bilayer graphene systems. (a) shows a twisted bilayer that hosts topologically \emph{trivial} domain walls. The walls that meet in the centre of the figure are shown by a dotted line. In (b) the upper layer has been uni-axially strained, thus allowing a set of equally spaced vertical \emph{topological} domain walls. Sample (c) contains a single line defect, where a line of blue atoms has been removed from the top layer, resulting in a step-like change of mass along the \emph{topological} domain wall.}
\label{figdws2}
\end{figure}

Having established the projection principle for the presence of a number of surface states, we now introduce the notion of an even-odd effect for domain wall states: \emph{A coupled $2n - 1$ topological domain-wall system has at least one single zero-energy mode localized on the domain walls.} This definition can be seen to be true from a simple bandstructure argument. For $2n-1$ domain walls, the mass changes sign from, say, positive to negative $n$ times, and negative to positive $n-1$ times. This leads to $n$ solutions at the $K$ point with positive slope, and $n-1$ solutions at the $K$ point with negative slope. When these solutions are admixed, we must do so pair-wise. Each pair becomes gapped, and we are left with a single zero energy solution at the $K$ point with positive slope. Similarly we are left with a single zero energy solution at the $K'$ point with negative slope.

Further, we can quantify this argument as follows. Consider a system of $N_W$ domain walls described by Eq. (\ref{hlow}), where the distance between neighbouring walls is $d$, and the overlap integral of these nearest neighbors is $t_D$. The Hamiltonian around the point $k_x = K^i_x$ is given by

\begin{equation}
H =  \sum_j^{N
_W} v_F(k_x-K^i_x) c_{k_x,j}^\dag c_{k_x,j} + \sum_j^{N_W} t_D(c_{k_x,j+1}^\dag c_{k_x,j} + \,\mathrm{c.c.})
\label{hlowp}
\end{equation}
where $t_D$ is the overlap integral of states on neighboring walls. In this case, the solution to the inter-domain wall coupling in the Hamiltonian takes the tri-diagonal form

\begin{equation}
H'-\mathrm{diag}_N(\epsilon) \rightarrow 
\begin{pmatrix}
-\epsilon & t_D & & \\
t_D & -\epsilon & t_D & \cdots\\
0 & t_D & -\epsilon & \\
 & \vdots &   & \ddots\\
\end{pmatrix}.
\label{matrix}
\end{equation}
Searching for $\epsilon = 0$ solutions, we find that only \emph{odd} dimensional matrices support such solutions. This confirms the \emph{even-odd} effect. We now turn to the discussion of lattice-mismatched bilayer graphene.

It has been shown recently that twisted bilayer graphene displaying a Moire pattern has an oscillating effective mass \cite{blgmoire}. This is a direct manifestation of the effect introduced in Fig. \ref{figmismatch}(a). In order to develop some intuition for this, consider two single, uncoupled layers of graphene. If we introduce only the dominant interlayer tunnelling term $\gamma$, then in the normal Bernal-stacked regime, $\gamma$ directly couples the $A$ sublattice of the upper species with the $B$ sublattice of the lower, as the upper $A$ site sits directly above the $B$ site. However, if the A-sublattices constitute interlayer nearest-neighbours, then $\gamma$ couples them instead. If we construct a single-layer effective theory by integrating out one of the two layers, then this interlayer coupling becomes an effective mass. The effective Hamiltonian is then none other than Eq. (\ref{H0}). We should emphasize that lattice-mismatching of single layer graphene with a substrate is in principle completely equivalent.

\begin{figure}[tbp]
\centering\includegraphics[width=8cm]{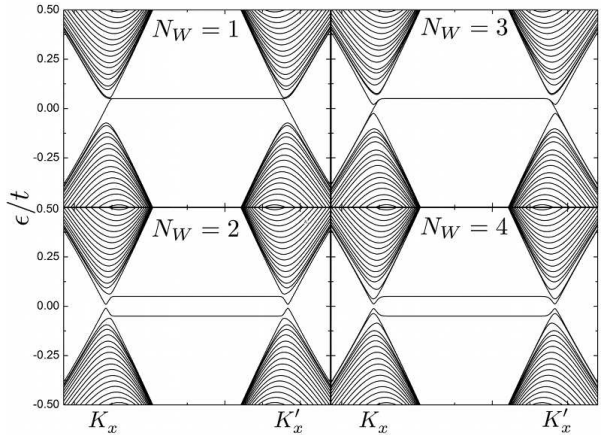}
\caption{Explicit bandstructure calculation demonstrating the even-odd effect in a typical zig-zag graphene nanoribbon under an oscillating staggered sublattice potential. $N_W$ is the number of domain walls in the sample, and we clearly see that when $N_W/2\in \mathbb{Z}$ only, the system is gapless.}
\label{figoddeven}
\end{figure}

The pattern of domain walls in a twisted bilayer exhibiting a Moire-like pattern is shown in Fig.  \ref{figdws2}(a). We have numerically confirmed that in the most promising situation
of small rotation angles, the domain walls
are oriented along the armchair direction. Two systems with `topological' walls are shown in Fig. \ref{figdws2}(b) and (c). The former is a set of parallel domain walls, obtained by uni-axial straining of the upper layer. This results in a continuously varying mismatch between the two lattices, and thus again a continuously varying, oscillating mass term. Fig.  \ref{figdws2}(c) shows a bilayer system where the upper layer has a single line defect. A line defect is obtained by removing a single line of A or B atoms, thus translating the entire lattice a small constant amount. Rather than a continuous variation of mass, this system contains a sharp step-like change in sign of the mass. For systems (a) and (b), the number of domain walls in the sample is tunable by either varying the angle of rotation or the amount of uni-axial straining, respectively. 

Finally, we confirm numerically the existence of a single zero-energy state in systems with an odd number of topological domain walls. In Fig. \ref{figoddeven} we show the dispersions between two projected $K$ points for typical zig-zag nanoribbons with $N_W = 1,2,3$ and $4$ domain walls.We find that in the presence of an odd number of topological domain
walls only a single domain-wall mode is gapless and all the other
modes have an energy gap which is determined by the tunneling energy.
This energy gap can be significant if the domain walls are close to
each other. We note that we have also confirmed the triviality of armchair domain wall systems, thus demonstrating the generalized projection principle.

In conclusion, we have shown by general topological arguments, that topologically trivial insulators with multiple massive Dirac cones, where a local change in mass induces a domain wall, support robust topological states. By a generalized `projection principle', we showed that the topological regime is that where the projection of the Dirac cones onto the domain wall does not cause the two cones to overlap. We also defined an even-odd effect for these systems, which demands that a system with an odd number of coupled domain walls necessarily supports at least a single zero energy mode. Mismatched bilayer graphene was then shown to be an ideal system to experimentally realize such states.

\end{document}